\documentclass[aps,prl,reprint,superscriptaddress,twocolumn,showkeys,amsmath,amssymb, longbibliography]{revtex4-1}
\usepackage[english]{babel}
\usepackage{amsmath,amssymb,bbm,mathrsfs,bm,braket,color,graphicx,comment,amsfonts,dsfont}
\usepackage[colorlinks,linkcolor=blue,citecolor=blue,urlcolor=blue]{hyperref}
\usepackage[mathscr]{euscript}
\usepackage{physics}
\usepackage{xcolor}
\usepackage{bm}
\usepackage{orcidlink}
\usepackage{multirow}

\usepackage{pifont}
\newcommand{\cmark}{\ding{51}}%
\newcommand{\xmark}{\ding{55}}%

\makeatletter
\def\maketitle{
\@author@finish
\title@column\titleblock@produce
\suppressfloats[t]}
\makeatother

\begin{document}

\title{
Physical properties of an Aperiodic monotile: \\
Graphene-like features, chirality and zero-modes
}
\author{Justin Schirmann
\orcidlink{0009-0007-7030-0155}
}
\email{justin.schirmann@neel.cnrs.fr}
\thanks{\\ These two authors contributed equally to this work.}
\affiliation{Univ. Grenoble Alpes, CNRS, Grenoble INP, Institut Néel, 38000 Grenoble, France}
\author{Selma Franca
\orcidlink{0000-0002-0584-2202}
}
\email{selma.franca@neel.cnrs.fr}
\thanks{\\ These two authors contributed equally to this work.}
\affiliation{Univ. Grenoble Alpes, CNRS, Grenoble INP, Institut Néel, 38000 Grenoble, France}
\author{Felix Flicker 
\orcidlink{0000-0002-8362-1384}}
\email{flicker@cardiff.ac.uk}
\affiliation{School of Physics and Astronomy, Cardiff University,
The Parade, Cardiff CF24 3AA, United Kingdom}
\affiliation{School of Physics, H.~H.~Wills Physics Laboratory, Tyndall Avenue, Bristol, BS8 1TL, United Kingdom}
\author{Adolfo G. Grushin\orcidlink{0000-0001-7678-7100}}
\email{adolfo.grushin@neel.cnrs.fr}
\affiliation{Univ. Grenoble Alpes, CNRS, Grenoble INP, Institut Néel, 38000 Grenoble, France}

\date{\today}

\begin{abstract}
The discovery of the Hat, an aperiodic monotile, has revealed novel mathematical aspects of aperiodic tilings. 
However, the physics of particles propagating in such a setting remains unexplored. In this work we study spectral and transport properties of a tight-binding model defined on the Hat. 
We find that (i) the spectral function displays striking similarities to that of graphene, including six-fold symmetry and Dirac-like features; (ii) unlike graphene, the monotile spectral function is chiral, differing for its two enantiomers; (iii) the spectrum has a macroscopic number of degenerate states at zero energy;  (iv) when the magnetic flux per plaquette ($\phi$) is half of the flux quantum, zero-modes are found localized around the reflected `anti-hats'; and (v) its Hofstadter spectrum is periodic in $\phi$, unlike for other quasicrystals. 
Our work serves as a basis to study wave and electron propagation in possible experimental realizations of the Hat, which we suggest.
\end{abstract}

\maketitle

\footnotetext[5]{By quasicrystals we mean structures that are aperiodic but long-range ordered, evidenced by a structure factor with well defined (Bragg) peaks. Alternatively we could refer to the Hat as an incommensurate crystal, as its structure factor displays crystalline sixfold symmetry.}
\paragraph{Introduction} --- Quasicrystals ~\cite{Shechtman:1984kf,Note5} exhibit a rich variety of physical properties beyond those observed in periodic crystals~\cite{Janot1992,Stadnik1999,Janssen2008}. 
While long-range ordered like crystals, they lack periodicity, leading to novel electronic~\cite{Kohmoto1986, Ping1987, Smith1987, Fujiwara1988, Tsunetsugu1990, Rieth98}, optical~\cite{Mayou2000, BURKOV1994525, Timusk2013}, vibrational~\cite{Janot1992, Janssen2008}, or topological phenomena~\cite{Arai1987, Hatakeyama1989, Piechon1999, Vidal2004, Kraus:2012iqa, Tran:2015cj, Fulga:2016jo, Fuchs:2016hp, Fuchs:2018dd, Huang2018, Huang2018b, He2019, Loring2019, Chen:2019vy, varjas_topological_2019, Huang:2020ho, Hua2020, Cain2020, Duncan2020, Zilberberg:21, Else2021, Fan2021,Hua2021, Ghadimi2021, Peng2021, Shi2022, Johnstone2022, Wang2022, Jeon2022}.  
Quasicrystalline materials derive their exotic behaviors from the symmetries of their quasilattices \footnote{Quasilattices are sets of delta functions with the symmetries of the physical quasicrystals, in exact analogy to the relationship between crystal lattices and crystals.}.
In two-dimensions (2D) quasilattice symmetries can often be described by aperiodic tilings of the plane~\cite{Janot1992,Janssen2008,Jaric1989,Levitov1988,Mermin1992}.

Recently Smith \textit{et al.}~\cite{Smith2023a, Smith2023b} discovered the first example of a single, simply connected tile that tiles the plane only aperiodically. 
Dubbed `the hat', the shape admits a continuous range of deformations with the same property. 
As with certain other quasicrystals, the tiling can be understood as a slice through a higher-dimensional periodic lattice~\cite{Socolar2023,Baake2023}. 
The Hat quasilattice (we use capitalization to distinguish the \textit{hat tile} from the \textit{Hat tiling}) is chiral; it has two enantiomers related by mirror symmetry. 
The tiles are two mirrored images of the same tile (Fig.~\ref{fig:spectral}(a)): the hat, colored white, and the anti-hat, colored blue. 
A related tile, tile(1,1), does not require its mirror image to tile the plane aperiodically~\cite{Smith2023b}.

Does the Hat imprint any novel physical properties on propagating particles compared to other two-dimensional aperiodic lattices? 
A fruitful strategy to answer this question is to define a vertex tight-binding model~\cite{Kohmoto1986, Fujiwara1988, Arai1988, Piechon1999} on the quasilattice, in which particles hop between nearest-neighbor vertices with equal probability. 
Vertex models have a single energy scale, the hopping, and hence conveniently isolate the effect of the graph connectivity on particle motion. 
They reveal unique spectral properties of quasicrystals, such as multifractal spectra~\cite{Fujiwara1988, Milde97, Rieth98, Piechon1999, Stadnik1999, Nicolas2017, Goblot2020, Jeon2022, Jagannathan2023, Jagannathan2023}, characteristic of critical disorder systems~\cite{evers_anderson_2008}, or exact zero-modes~\cite{Kohmoto1986, Arai1988, Fujiwara1988, Flicker2020, Oktel2021}. 
Vertex models on 2D quasilattices differ from those of periodic 2D lattices by displaying an aperiodic Hofstadter spectrum as a function of an applied perpendicular magnetic flux per plaquette $\phi$~\cite{Arai1987, Hatakeyama1989, Piechon1999, Vidal2004, Tran:2015cj, Fuchs:2016hp, Fuchs:2018dd, Johnstone2022, Jeon2022}.  

In this work, we establish the spectral and transport properties of the Hat through its vertex tight-binding model. 
The momentum-resolved spectral function displays striking similarities with that of graphene, including putative Dirac cones and six-fold symmetry. 
However, unlike graphene, the Hat's spectral function is chiral, and displays a predictable finite density of exact zero-modes.
As the Hat is a monotile, its Hofstadter spectrum is periodic in magnetic field flux, bypassing incommensurability effects of `polytiled' quasicrystals~\cite{Arai1987,Hatakeyama1989,Piechon1999,Vidal2004,Tran:2015cj,Fuchs:2016hp,Fuchs:2018dd,Johnstone2022}.
The Hofstadter bands carry a Chern number that quantizes the two-terminal conductance in units of $e^2/h$.
Hence, the physical properties of the hat introduce a remarkable new class of phenomena between periodic crystals and aperiodic quasicrystals.

\begin{figure*}
    \centering
    \includegraphics[width=\linewidth]{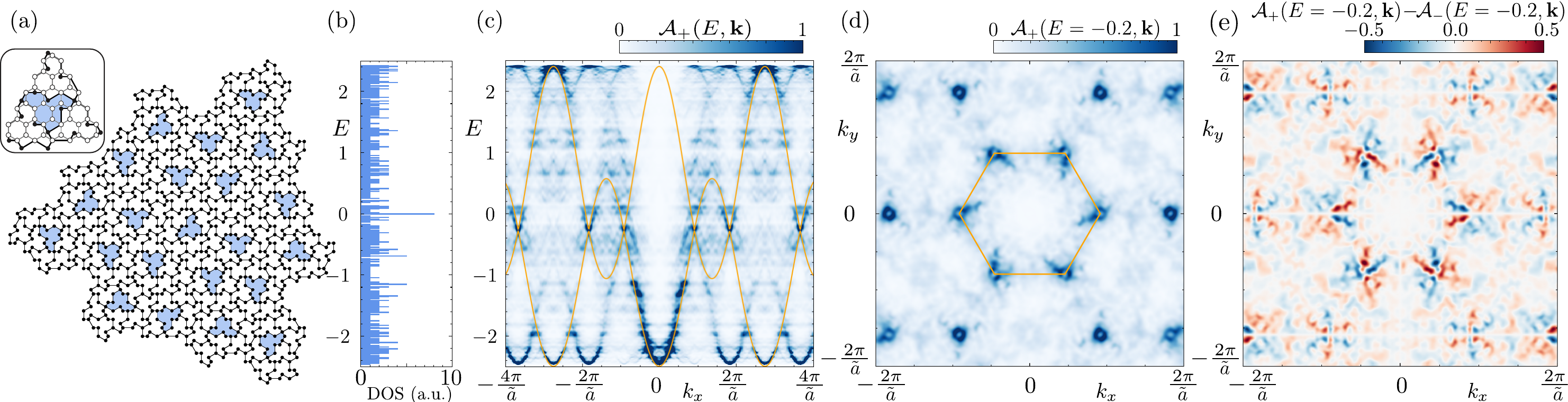}
    \caption{\label{fig:spectral} {Spectral properties of the vertex model Eq.~\eqref{eq:Ham} on the Hat.} 
    (a) H2, with $22$ anti-hats (reflected images of the hat tile) colored in blue. The inset shows a patch of the Hat quasilattice (solid circles) overlayed with a graphene hexagonal lattice approximant (open circles).
    (b) Density of states of the vertex model Eq.~\eqref{eq:Ham} on H2. 
    (c) Momentum-resolved spectral function $\mathcal{A}_+(E,\mathbf{\mathbf{k}})$ of the enantiomer in (a) along the ${k}_x$ momentum direction, calculated using the Kernel Polynomial method~\cite{weise_kernel_2006}. The dispersion relation Eq.~\eqref{eq:Graph} for the hexagonal lattice (inset of (a)), Eq.~\eqref{eq:Graph}, is overlaid in orange, with parameters $(a_g,t_1,t_2,\varepsilon_0) = (2a/\sqrt{3}, 0.82 , -0.025, -0.2)$. 
    (d) $\mathcal{A}_+(E=-0.2,\mathbf{k})$ as a function of momentum $\mathbf{k}=(k_x, k_y)$. 
    (e) Difference between the spectral functions 
    $\mathcal{A}_\pm(E=-0.2,\mathbf{\mathbf{k}})$ of the system in (a) and its reflection in the \textit{y}-axis.
}
\end{figure*}

\paragraph{Spectral properties} --- The Hat quasilattice can be generated using `inflation rules' in which four basic metatiles (combinations of hats), dubbed H, P, T and F, divide into smaller versions of the same tiles, which are then inflated (rescaled) so all hats return to their original size~\cite{Smith2023a} (the Supplemental Material (SM)~\cite{SuppMat} contains a brief introduction).
We define a tight-binding model on the vertices of inflations of all metatiles.
We focus on H metatiles, whose second inflation, H2, is shown in Fig.~\ref{fig:spectral}(a).
Unless stated otherwise, our results apply to all tilings of the same hat tile.

Each vertex is either two-, three- or four-fold coordinated, separated by three possible bond lengths. 
The average coordination number is $\left\langle z \right\rangle \sim 2.31$ and the average bond length is $ \Tilde{a} = 1.37 a$, with $a$ the shortest bond length.

Setting all hoppings equal defines the vertex Hamiltonian~\cite{Kohmoto1986,Fujiwara1988,Arai1988}
\begin{equation} \label{eq:Ham}
    H_{\mathrm{Hat}} = -t \sum_{\left\langle ij\right\rangle} c^{\dagger}_{i}c_{j} +\mathrm{h.c}.
\end{equation} 
The sum runs over all pairs of neighboring sites and the operators $c^{\dagger}_i$ and $c_i$ create and annihilate a particle on site $i$. 
We choose $t=1$ without loss of generality.

The density of states (DOS) is shown in Fig.~\ref{fig:spectral}(b). 
The energy minimum $E_m \approx -2.4$ is well captured by the average coordination number $\left\langle z \right\rangle$~\cite{Kohmoto1986, Fuchs:2018dd}. 
Like other quasicrystals~\cite{BURKOV1994525, Rieth98}, the Hat exhibits a fractal DOS with a multitude of van Hove singularities. 

The probability of finding a state at energy $E$ and momentum $\mathbf{k}$ is determined by the spectral function, $\mathcal{A}(E,\mathbf{k})=\bra{\mathbf{k}}\delta(H_{\mathrm{hat}}-E)\ket{\mathbf{k}}$, shown in Figs.~\ref{fig:spectral}(c) and (d) for H2.
This function is well defined even without translational invariance~\cite{marsal_topological_2020, Marsal2022, corbae2023, Ciocys2023, Rotenberg2000, Rotemberg2004, Rogalev2015} as it measures the overlap of the eigenstates with plane-waves of well-defined momentum $\mathbf{k}$, $\left\langle{\mathbf{r}}|{\mathbf{k}}\right\rangle=\frac{1}{\sqrt{N}}e^{\iota\mathbf{k}\cdot\mathbf{r}}$, with $\iota^2 = -1$, $N$ the number of vertices and $\mathbf{r}=(x,y)$ their positions.
The spectral function has been measured using angle-resolved photoemission experiments in quasicrystals~\cite{Rotenberg2000, Rotemberg2004, Rogalev2015}.

The spectral function shows, close to zero energy, Dirac node-like features reminiscent of graphene's band structure.
To quantify this similarity we define a periodic hexagonal lattice, shown as empty circles in the inset of Fig.~\ref{fig:spectral}(a), that we call the graphene approximant.
The graphene lattice constant $a_g=2 a/\sqrt{3}$ is chosen such that the associated honeycomb lattice captures many ($\approx 53\%$) of the Hat's vertices. 
The graphene approximant's dispersion and periodicity capture several features of $\mathcal{A}(E,\mathbf{k})$ (orange lines in Figs.~\ref{fig:spectral}(c), (d)). 
The dispersion relation is~\cite{CastroNeto2009}
\begin{equation}\label{eq:Graph}
    E_{\pm}(\mathbf{k}) = \pm t_1 \sqrt{3+f(\mathbf{k})}-t_2 f(\mathbf{k})+\varepsilon_0,
\end{equation}
where $f(\mathbf{k})= 2\cos(\sqrt{3}k_x)+4\cos(\sqrt{3}k_x/2)\cos(3k_y/2)$, $t_1$ and $t_2$ are the first and second nearest-neighbor hopping amplitudes and $\epsilon_0$ is the energy offset. 
Here, we use $(a_g,t_1,t_2,\varepsilon_0) = (2 a/\sqrt{3}, 0.82 , -0.025, -0.2)$.
A nonzero $t_2$ and $\epsilon_0$ mimic the Hat spectrum's asymmetry.
The Brillouin zone of the graphene approximant is represented with orange lines in Fig. \ref{fig:spectral}(d). 
The zone corners match the location of the Dirac node-like features close to $E=-0.2$. 
The Hat's $C_6$ symmetry~\cite{Socolar2023} is apparent in Fig.~\ref{fig:spectral}(d).

Unlike graphene, we can define an enantiomer of the Hat quasilattice by applying a reflection operator.
The difference between the two enantiomorphic spectral functions 
$\mathcal{A}_{+}(E,\mathbf{k})-\mathcal{A}_{-}(E,\mathbf{k})$ at $E=-0.2$ reveals chiral properties spread across the quasi-Brillouin zone (Fig. \ref{fig:spectral}(e)).

\begin{figure}
    \centering
    \includegraphics[width=\columnwidth]{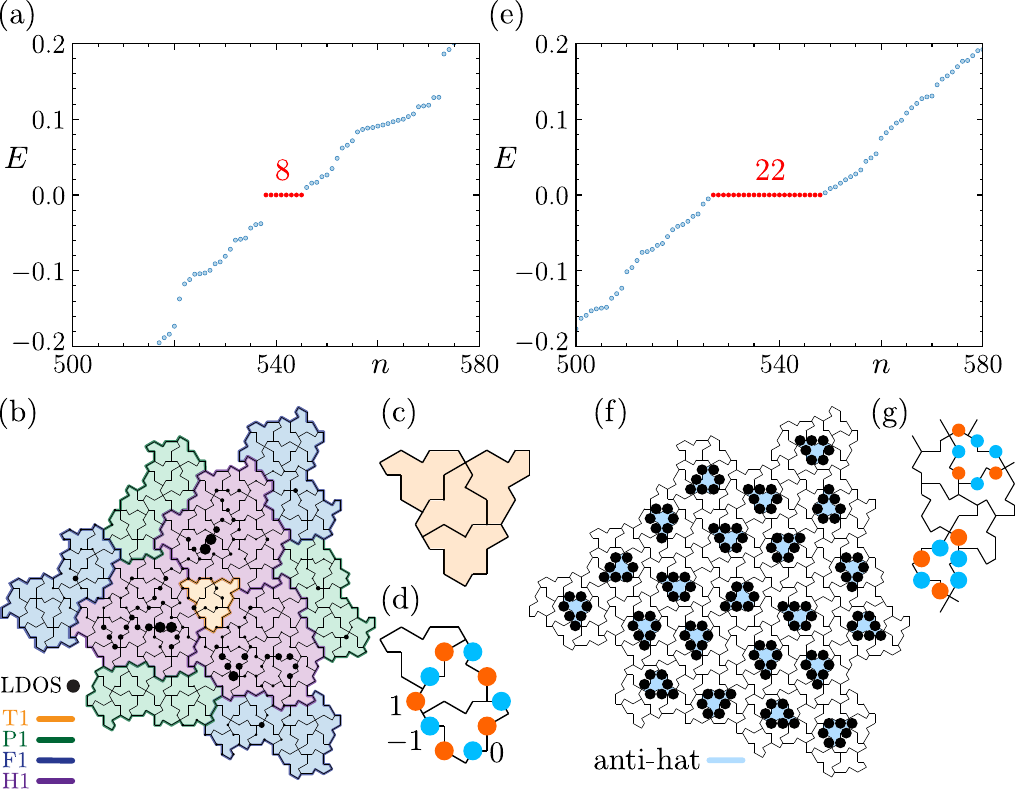}
    \caption{\label{fig:zeromode} {Zero-modes under $0$- and $\pi$-flux.}
    (a) Low-energy spectrum of the Hat quasilattice without flux. The eight exact zero-modes are colored in red. 
    (b) The associated local density of states (LDOS) of these zero-modes. Colors highlight previous inflation generations (T1, P1, F1 and H1) composing H2.
    (c) The T1 quasicrystallite has no zero-modes.
    (d) The T1 quasicrystallite without the rightmost hat has a single zero-mode. The overlaid zero-mode amplitudes form the Sutherland loop sequence $\{0,1,0,-1\}^m$\cite{Sutherland1986} of length $4m$, with $m\in\mathbb{Z}$.
     (e) The low-energy spectrum under $\pi$-flux. Red: the 22 exact zero-modes.
     (f) Corresponding zero-energy LDOS, pinned to anti-hats.
     (g) Zoom of the wave-function corresponding to eigenstate $530$. Because $\phi/\phi_0=1/2$, the Sutherland loop is modified to have one defect per anti-hat.
}
\end{figure}

\paragraph{Zero-energy states} --- Another striking feature dissimilar to graphene is the existence of a finite density of zero-energy states (Fig.~\ref{fig:zeromode}(a)).
The number of these zero-modes can increase or decrease upon adding tiles (see SM~\cite{SuppMat} for examples). 
Similar zero-modes have been found in numerous quasilattices, including the Penrose tiling~\cite{Kohmoto1986,Day-Roberts2020,Flicker2020,Bhola2022}, the Ammann-Beenker tiling~\cite{Oktel2021,Koga17,Lloyd2022}, and quasicrystalline graphene bilayers~\cite{Ha2021}. 
Fig.~\ref{fig:zeromode}(b) shows the local density of states (LDOS) at zero energy of H2.
While some of the zero-mode weight arises from zero-modes of underlying metatiles of the previous inflation generation (colored areas in Fig.~\ref{fig:zeromode}(b)), the finite weight on T1 within H2, Fig.~\ref{fig:zeromode}(b), is absent in T1 itself (Fig.~\ref{fig:zeromode}(c)).

However, all these zero-modes can be understood in terms of the graph connectivity. The tight-binding model defines the adjacency matrix of the corresponding graph; the zero-modes then span its null space. A basis can always be found in which each (un-normalized) zero-mode has integer amplitudes on all vertices (see SM~\cite{SuppMat} for proof). These modes are fragile in the sense that they rely on equal hoppings to remain at strictly zero energy~\cite{Lloyd2022,Bhola2022}.

For a nearest-neighbor vertex model, like Eq.~\eqref{eq:Ham}, the zero energy condition implies that for every site $i$, the sum of the amplitudes on all neighbors $j$ of $i$ must vanish: $\sum_{j} \Psi_j =0$~\cite{Sutherland1986}. 
In the Hat tiling the simplest zero-modes take the form of cycles of length $4m$, with $m$ integer. Here the amplitudes around the cycle can be taken to be the repeated sequence $\{0,1,0,-1\}^m$ whenever vertices from the rest of the graph connect only to cycle vertices of zero amplitude or connect to pairs of vertices with opposite amplitude. This form is a generalization of one identified by Sutherland~\cite{Kohmoto1986,Sutherland1986}; we term it a Sutherland loop.

In Figs.~\ref{fig:zeromode}(c) and (d) we show two quasicrystallites differing by one hat. 
We find no zero-modes in (c) and one in (d). 
Removing the extra hat allows for a 20-vertex Sutherland loop shown in Fig.~\ref{fig:zeromode}(d).
Larger quasicrystallites contain integer-amplitude zero-modes not of the Sutherland loop form. 
In general, all integer-amplitude zero-modes of a given quasicrystallite can be exactly enumerated using the {\it Hermite normal form} of the adjacency matrix (see SM~\cite{SuppMat})~\cite{bookCohen,Kannan1979}.

\paragraph{$\pi$-flux zero-modes} --- The shortest loops, circling a single tile, cannot be Sutherland loops as they have a length of 13 or 14.
However, anti-hats, the enantiomorphic minority tiles which have two four-fold coordinated sites (blue in Fig.~\ref{fig:spectral}(a)), can support a zero-mode if we allow a sign flip of one of the hoppings around the anti-hat -- equivalent to threading a magnetic flux of $\pi$ per anti-hat.
As all the hats and anti-hats have the same area, this suggests that applying a perpendicular magnetic field with exactly $\pi$-flux per plaquette should generate one zero-mode per anti-hat. 
In fact, we find that these are the only zero-modes in this setting.

To model a perpendicular magnetic field $B$ we introduce a Peierls phase by changing the hopping from site $j$ to site $i$ as $t\to t\ \mathrm{exp}(-\iota  \pi \frac{\phi}{\phi_0}(x_i-x_j)(y_i+y_j)/\mathcal{A})$,
where $\mathcal{A}=8\sqrt{3} a^2$ is the hat area, $\phi_0=h/e$ is the magnetic flux quantum, and $\phi$ is the magnetic flux. 
When $\phi/\phi_0=1/2$ the hoppings can be chosen to be real, with every hat tile having an odd number of negative bonds compared to Eq.~\eqref{eq:Ham}. 
The spectrum close to $E=0$ is shown in Fig.~\ref{fig:zeromode}(e). 
For inflations of primitive metatiles H0, T0, P0 and F0 that we checked, we found that the number of zero-modes equals the number of anti-hats (see SM \cite{SuppMat}).
Their LDOS is localized exactly at the anti-hats, as exemplified by Fig.~\ref{fig:zeromode}(f).
Furthermore, Fig.~\ref{fig:zeromode}(g) shows that the zero-mode wave-function, having one defect per anti-hat, disobeys the Sutherland loop form.

\paragraph{Hofstadter spectrum} --- Fig.~\ref{fig:magn}(a) shows the bulk spectrum of H3, the third inflation of the H metatile, 
as a function of $\phi/\phi_0$.
To isolate the bulk spectrum we exclude states whose weight inside the yellow line in Fig.~\ref{fig:magn}(b) is smaller than their weight outside it.

The spectrum is periodic in the interval $\phi/\phi_0\ \in [0,1)$, as there is only one type of plaquette and only one area $\mathcal{A}$ to normalize the flux. 
This is a special feature of monotiles compared to other quasicrystals.
A typical quasicrystal spectrum is aperiodic as a function of $\phi/\phi_0$~\cite{Hatakeyama1989,Duncan2020,Ghadimi2022,Johnstone2022}, excepting quasicrystals composed of tiles with commensurate areas~\cite{Fuchs:2016hp,Fuchs:2018dd}.

When the magnetic length $l_B \gg a$, the spectrum splits into Landau levels near the bottom of the spectrum ($E \sim - 2.4 $), and disperses linearly with $B$~\cite{Fuchs:2018dd}.
In the Hofstadter regime $l_B \sim a$, the spectrum is split into Hofstadter bands separated by gaps.
Placing the Fermi energy $E_F$ within these gaps, the system should display topological edge states and a quantized Hall conductance~\cite{thouless1982, niu1985}.

The topological properties of aperiodic 2D systems without time-reversal symmetry
are captured by a quantized bulk average $\mathcal{C}$ of the local Chern marker $\mathcal{C}_{\mathbf{r}}$~\cite{bianco11}.
Mathematically, $\mathcal{C} = \frac{1}{\mathcal{A}_b} \sum_{\mathbf{r} \in \mathcal{A}_{b}} \mathcal{C}_{\mathbf{r}}$, with $\mathcal{A}_b$ the area of a bulk region highlighted in Fig.~\ref{fig:magn}(b), $\mathcal{C}_{\mathbf{r}} = \bra{\mathbf{r}}\hat{C}\ket{\mathbf{r}}$ and
\begin{equation}
    \hat{C} = 2 \pi \iota \left(\hat{\mathcal{P}}\hat{X}\hat{\mathcal{Q}}\hat{Y}\hat{\mathcal{P}}-\hat{\mathcal{P}}\hat{Y}\hat{\mathcal{Q}}\hat{X}\hat{\mathcal{P}}\right).
\end{equation}
Here $\hat{\mathcal{P}} = \sum_{E<E_F}\ket{\Psi}\bra{\Psi}$ is the projector onto occupied states, $\hat{\mathcal{Q}} = 1 - \hat{\mathcal{P}}$, and $\hat{X}$ and $\hat{Y}$ are position operators. 
In a topological phase $\mathcal{C}\in \mathbb{Z}$ coincides with the Chern number for periodic systems~\cite{bianco11}.

Fig.~\ref{fig:magn}(b) shows $\mathcal{C}_{\mathbf{r}}$ for $E_F= -1.58$ and $\phi/\phi_0 = 0.2$; $\mathcal{C}\approx -1$ in the bulk, as expected for a topologically non-trivial phase~\cite{bianco11}. 
The bulk value of $\mathcal{C}$ in the interval $\phi/\phi_0 \in [0,1/2]$ is shown in Fig.~\ref{fig:magn}(c).
Regions with nontrivial Chern numbers match the positions of bulk gaps in Fig.~\ref{fig:magn}(a), confirming the presence of topological edge states.

The physical imprint of topological edge states is a quantized two-terminal conductance $G$, in units of $G_0= e^2/h$.
To confirm this, we attach leads to two boundary regions of H3 (Fig.~\ref{fig:magn}(b)).
These leads consist of decoupled waveguides of dispersion $-2 \cos{k_z}$, oriented along the \textit{z}-direction and placed such that each waveguide probes a single site.
The two-terminal conductance $G$ is then defined in the Landauer-Buttiker formalism~\cite{datta_2005} as $G = G_0 \rm Tr[\tau \tau^{\dagger}$], where $G_0=e^2/h$ is the conductance quantum and $\tau$ is the transmission matrix between two leads calculated using \textsc{Kwant}~\cite{groth_kwant_2014}.

The conductance map as a function of $\phi/\phi_0$ and $E$ is shown in Fig.~\ref{fig:magn}(d).
In regions with non-zero density of bulk states, the conductance exhibits fluctuations observed in topologically trivial quasicrystals~\cite{Stadnik1999}.
However, inside the bulk gaps with nontrivial $\mathcal{C}=\pm 1$, $G = G_0$, stemming from a topological boundary state.
Finite size effects are visible in smaller gaps with larger $\mathcal{C}$ (Fig.~\ref{fig:magn}(c)), where $G$ deviates from quantization.

\begin{figure}
    \centering
    \includegraphics[width=\columnwidth]{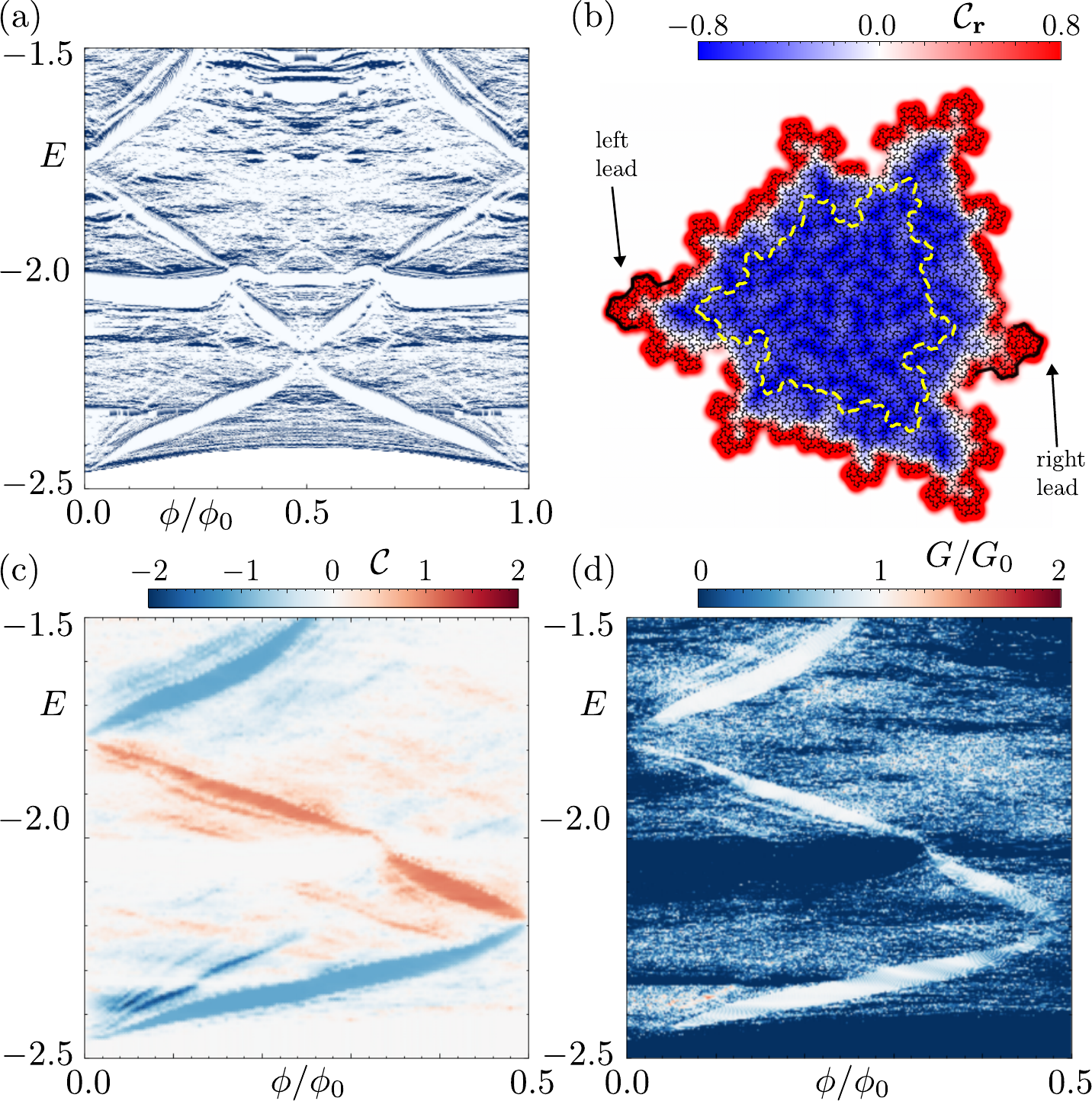}
    \caption{\label{fig:magn} {Hofstadter spectrum and quantized conductance.}
    (a) The bulk Hofstadter spectrum Eq.~\eqref{eq:Ham} calculated for H3 near the band bottom. 
    (b) The local Chern marker $\mathcal{C}_{\mathbf{r}}$ calculated for H3 (black) at $E=-1.58$ and $\phi/\phi_0 = 0.2$.
    The yellow dashed line delimits the bulk area $\mathcal{A}_b$ used to produce panels (a) and (c). Inside $\mathcal{A}_b$, the marker averages to $\mathcal{C}=-1.03$.
    Thick black lines on the edges denote sites where leads are attached.
    (c)  $\mathcal{C}$ as a function of $E$ and  $\phi/\phi_0$. Bulk gap regions of panel (a) have nontrivial Chern numbers.
    (d) Corresponding two-terminal conductance map. Where $|\mathcal{C}| = 1$, the conductance is quantized to $G = G_0= e^2/h$, confirming the role of topological boundary modes in transport. The leads can be placed arbitrarily so long as they are well separated.}
\end{figure}

\paragraph{Physical properties of Tile(1,1)} --- With a geometric modification, the alternative monotile tile(1,1) tiles the plane only aperiodically without its mirror image~\cite{Smith2023a,Smith2023b}. 
This modified form is called `the spectre'; since we are interested only in graph connectivity we use the names Tile(1,1) and Spectre interchangeably. 
Tile(1,1) contains special tiles similar to the anti-hats.
They appear $\pi/6$ rotated from the other tiles, which appear only $\pi/3$ rotated from one another. 
As with anti-hats, these `anti-spectres' always have two four-fold coordinated vertices. 
As tile(1,1) is a member of the continuous family of tiles connected to the hat it is pertinent to ask which properties it maintains.

Unlike the Hat, the vertices of Tile(1,1) are not known to fit to a periodic hexagonal lattice. 
The graphene-like features are therefore washed out. However, the quasilattice remains chiral, so it remains true that $\mathcal{A}_{+}(E,\mathbf{k})-\mathcal{A}_{-}(E,\mathbf{k})\neq 0$.
Tile(1,1) also displays strictly localized zero-modes that have a similar origin to those in the Hat. 
In the presence of a $\pi$-flux, each anti-spectre again localizes a zero-mode. In all the cases we have checked, these again exhaust all zero-modes.
As the Spectre is a monotile, the Hofstadter spectrum is periodic in $\phi$
and supports topological gaps with quantized conductance, as for the Hat tiling.

\paragraph{Conclusions} --- The Hat and Spectre monotiles present physical properties that set them aside from previously known 2D crystals and quasicrystals. 
The Hat is unique as a quasicrystal in displaying graphene-like features in its spectral function, notably putative Dirac cones.
These are intrinsic, unlike in quasicrystalline graphene bilayers~\cite{Ahn_2018,Yu2019,Shi2022} where they are inherited from the underlying graphene layers.
In the Hat, the intrinsic Dirac-cone-like features resemble those found in graphene in the presence of a small density of topological defects~\cite{Kot2020}.
Both the Hat and Tile(1,1) exhibit zero-modes.
At zero field, their origin is similar to those found in the Penrose and the Ammann-Beenker tilings.
However, imposing half a flux quantum per plaquette brings interesting differences, localizing the spectral weight around special tiles. 
Comparing the precise localization properties of the zero modes found in different quasicrystals merits further study.
Lastly, the Hofstader butterfly is periodic for monotilings, unlike for other studied quasicrystals.
In short, the Hat and Tile(1,1) display a blend of crystallinity and quasicrystallinity that sets them apart from known crystals and quasicrystals.

A vertex model of the tiling may be realized in metamaterials, where complex phases mimicking magnetic fields can be engineered \cite{Bandres2016}. 
Quasicrystals have been realized in photonic metamaterials~\cite{Notomi2004, Levi2011, Kraus:2012iqa, Bandres2016}, polaritonic systems~\cite{Tanese2014}, electrical circuits~\cite{Stegmaier2023}, microwave networks~\cite{Vignolo2016}, and acoustic~\cite{Chen2020} and mechanical~\cite{Wang2020b} metamaterials. 

While no material has yet been discovered with the symmetries of the Hat, it would seem likely that nature would realize such an elegant construction, just as Penrose tilings were discovered to describe the surfaces of icosahedral quasicrystals~\cite{Penrose74, Levine84, Bursill1985, McGrathEA02}.
A promising solid-state platform is the engineered adsorption of atoms to constrain scattering of surface states.
CO molecules on metals have been used to construct artificial honeycomb~\cite{Gomes2012} and fractal~\cite{Kempkes2019} lattices. 
2D lattices of Shiba states caused by magnetic adatoms on superconductors serve to engineer topological phases~\cite{nadj2013proposal,Schneider2022,Soldini2023}.

Chiral crystals display richer physical responses than achiral crystals~\cite{Babaev2023}. 
The Hat should share some of these features with chiral crystalline counterparts, including magnetochiral anisotropy~\cite{Morimoto2016} and optical gyrotropy~\cite{Ma2015,Zhong2016,Flicker2018,Wang2020}.
Additionally, adding defects or interactions can generalize other works on interacting quasicrystals~\cite{Hu2011,Koga17,Duncan2020,Liu2022}.
We leave the study of these effects for subsequent work.

\paragraph{Acknowledgements} 

We thank E. K\"{o}nig for illuminating discussions. 
We also thank the authors of Refs.~\onlinecite{Smith2023a, Smith2023b}, especially Craig Kaplan for publishing their programs to generate Hat tilings~\cite{weblink}.
A.G.G. and S. F. acknowledge financial support from the European Union Horizon 2020 research and innovation program under grant agreement No. 829044 (SCHINES). J. S. is supported by the program QuanTEdu-France n° ANR-22-CMAS-0001 France 2030. F.F. was  supported by EPSRC grant EP/X012239/1.
A. G. G. is also supported by the European Research Council (ERC) Consolidator grant under grant agreement No. 101042707 (TOPOMORPH). 
This research was supported in part by the National Science Foundation under Grant No. NSF PHY-1748958.
 The code used to generate our results is available at Ref.~\cite{schirmann_2023_8215399}.

\bibliography{e-hat.bib}

\clearpage
\newpage

\title{Supplemental Material to:  Physical properties of an Aperiodic monotile: \\
Graphene-like features, chirality and zero-modes}

\maketitle

\setcounter{secnumdepth}{5}
\renewcommand{\theparagraph}{\bf \thesubsubsection.\arabic{paragraph}}

\renewcommand{\thefigure}{S\arabic{figure}}
\setcounter{figure}{0} 

\appendix


\section{Brief introduction to the Hat and Tile(1,1) tilings}

The discovery of the Hat solves a long standing question: is there a single tile that tiles the plane only aperiodically, under rotations and translations? Formerly, the minimal tile set with this property contained two tiles. The Penrose tiles were the first known example~\cite{Penrose74}.
The \textit{hat tile} is one possible choice among a continuous family of tiles parameterized by two non-negative real numbers $(a, b)$ which is made possible by the fact that there are two lengths involved and sides of same length come in parallel pairs.

\begin{figure}[h]
    \centering
    \includegraphics[width=\columnwidth]{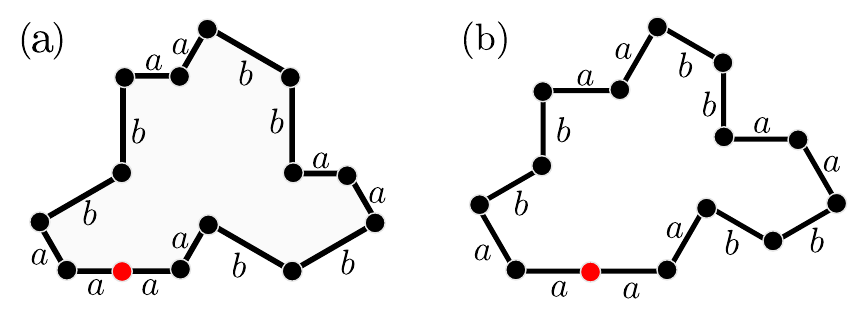}
    \caption{\label{fig:hatvstile} Hat and tile(1,1) (a) The hat belongs to an infinite family with of tiles with edge lengths $(a,b)$. The hat (a) corresponds to choosing $(a,b) = ( 1,\sqrt{3})$ while tile(1,1) (b) corresponds to the choice $(a,b) = (1,1)$. The red circle indicates an additional vertex that occurs at the intersection of some neighbouring tiles.}
\end{figure}

\begin{figure}[h]
    \centering
    \includegraphics[width=\columnwidth]{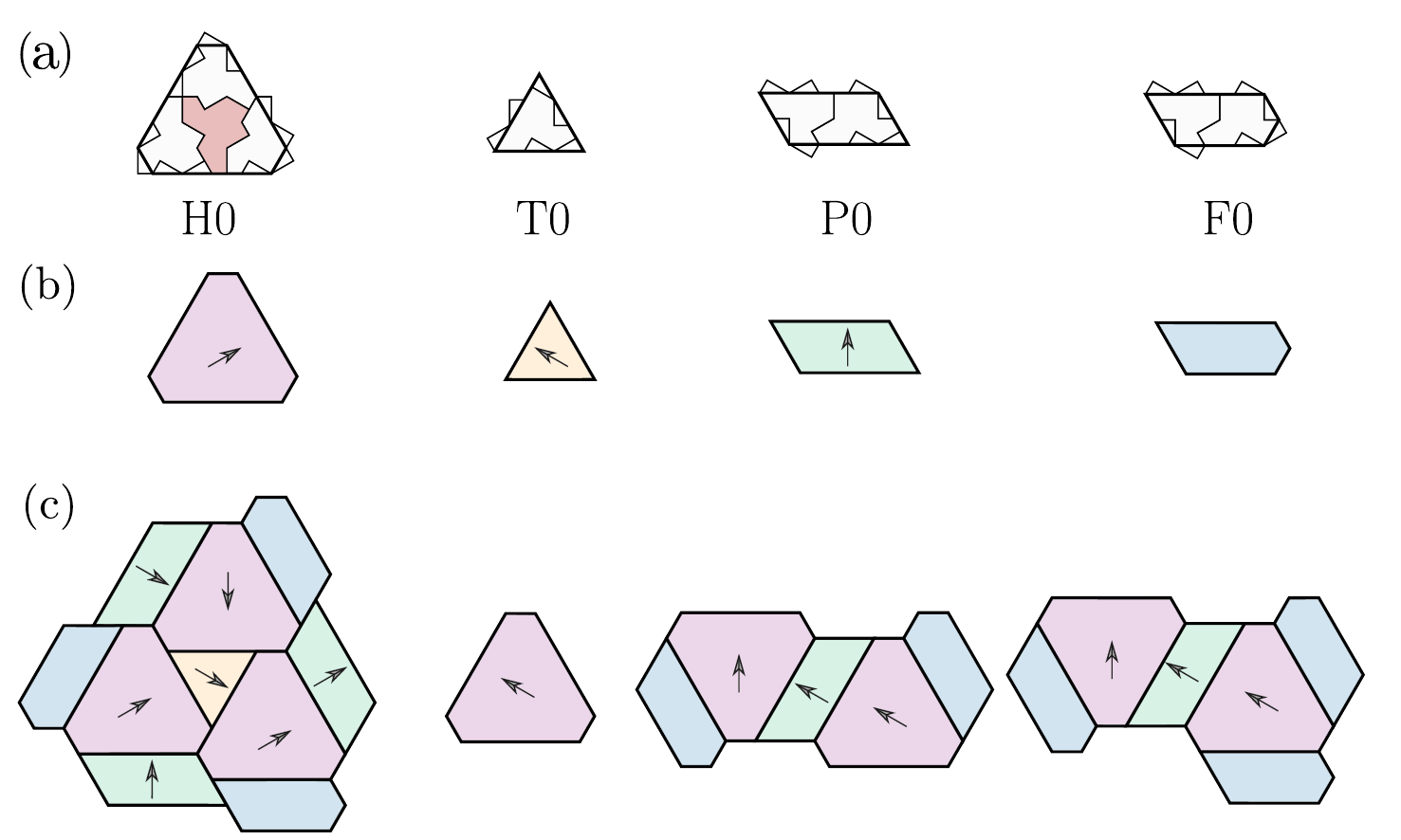}
    \caption{\label{fig:Metatiles} Inflation rules (a) The four initial metatiles (from left to right) H0, T0, P0 and F0 with $H_0=4, T_0=1, P_0=2$ and $F_0=2$ hats, respectively. The colored tile at the center of H0 corresponds to an anti-hat, all the others are regular hats. 
    (b) We assign to each of the metatiles an orientation. 
    (c) Inflation rules, at each step of the process, the metatiles are combined respecting the orientation constraints. Finaly, we replace the metatiles by the corresponding patches of hats.}
\end{figure}

The \textit{Hat tilling} is then obtained by defining a set of four metatiles and iterating the substitutions rules presented in Fig.~\ref{fig:Metatiles}.

The number of hats in each iteration is given by the following recursion equations:

\begin{eqnarray} \rm
H_{n+1} &=& 3H_{n}+T_{n}+3P_{n}+3F_{n},\\
T_{n+1} &=& H_{n},\\
P_{n+1} &=& 2H_{n} + P_{n} + 2F_{n},\\
F_{n+1} &=& 2H_{n} + P_{n} + 3F_{n},
\end{eqnarray}
where $ H_n$, $T_n$, $ P_n$ and $ F_n$  denote the number of hats contained in the metatile at step $n$. In the main text we present results for tiles H2 and H3, but all of our results generalize to any Hat tiling unless otherwise mentioned.

\begin{figure}[h]
    \centering
    \includegraphics[width=\columnwidth]{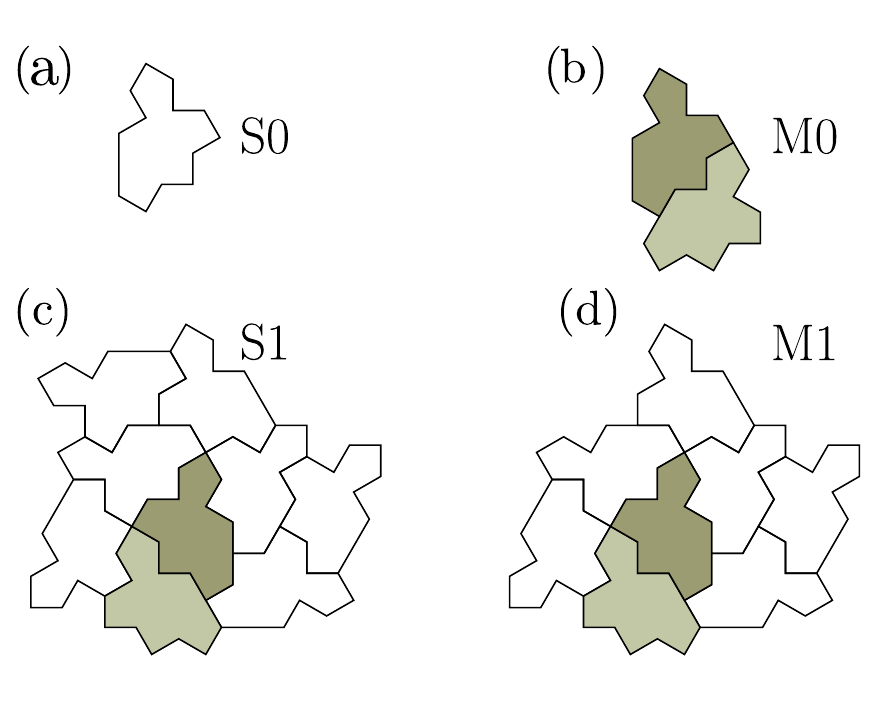}
    \caption{\label{fig:tile11} (a) Single compound made of Tile(1,1). (b) Two Tile(1,1) compound named Mystic. (c) Substitution rule for the Tile(1,1). (d) Substitution rule for the Mystic. It is important to note that at each step the whole tilling is mirrored. }
\end{figure}

Tile(1,1) tiling can be obtained through a similar process.
We denote two types of metatiles S and M (short for Mystic), where S0 is made of one tile(1,1) and M0 of two tiles(1,1), see Figs.~\ref{fig:tile11}(a) and (b).
Then, the number of tiles(1,1) in each iteration is given by the following recursion equations:
\begin{eqnarray} \rm
S_{n+1} &=& M_n + 7S_n., \\
M_{n+1} &=& M_n + 6S_n.
\end{eqnarray}

As a summary, we compare here the physical properties associated to the Hat and Tile(1,1) tilings discussed in the main text:
\begin{table}[h!]
    \centering
    \begin{tabular}{ |p{5.45cm}||p{1.25cm}|p{1.25cm}|  }
 \hline
 Properties & Hat & Tile(1,1)\\
 \hline
Tilling contains different enantiomers & \cmark & \xmark \\
Periodic Hofstadter butterfly  & \cmark   & \cmark  \\
Localized zero-modes ($\pi$-flux)  & \cmark   & \cmark  \\
Graphene-like features  & \cmark   & \xmark  \\
 \hline
\end{tabular} 
    \caption{Summary of the observed properties of the Hat and Tile(1,1).}
    \label{tab:Tile(11)Hats}
\end{table}

\section{Further details on zero-modes}

\subsection{Zero-modes at zero magnetic field}

In the main text we have discussed the conditions that the wave function must satisfy to be a zero-mode at $\phi/\phi_0=0$.
In Table~\ref{tab:zeroBmodes} we list the number of zero-modes for the different inflations of the primitive metatiles (H0,T0,P0 and F0).

\begin{table}[h!]
    \centering
    \begin{tabular}{ |p{2cm}||p{1cm}|p{1cm}|p{1cm}| p{1cm}|  }
 \hline
 Inflation & H & T & P & F\\
 \hline
0  & 0   & 0 & 1 & 1 \\
1  & 0   & 0 & 0 & 0 \\
2  & 8   & 0 & 1 & 1 \\
3  & 51   & 8 & 27 & 32 \\
 \hline
\end{tabular} 
    \caption{Number of exact zero-modes for each generation of inflation rules in the case of $\phi/\phi_0 = 0$.}
    \label{tab:zeroBmodes}
\end{table}

For completeness, we comment on the expected robustness of these zero-modes compared to other quasicrystalline zero-modes.
The zero-modes found on the Penrose tiling are topologically protected, in the sense that they survive arbitrary changes to the hopping integrals provided these remain nonzero.
This is a consequence of Lieb's theorem for bipartite graphs~\cite{Lieb89}, and can be understood in terms of dimer matchings~\cite{Flicker2020,Bhola2022,Lloyd2022}.
In the Ammann-Beenker tiling the zero-modes are fragile: they receive no such protection, and changing the hoppings breaks the local symmetries of the graph and lifts the degeneracy of these modes~\cite{Lloyd2022}. 
As the Hat tiling is non-bipartite, Lieb's theorem does not apply, and the zero-modes are expected to be fragile. 

\subsection{Zero-modes at $\pi$-flux}

Introducing a $\pi$-flux per plaquette, we have observed that the number of modes with energies smaller than $5 \times 10^{-5}$ matches the number of anti-hats in the Hat quasicrystallites of generations $1,2$ and $3$, as shown in Table~\ref{tab:piBmodes}.

\begin{figure}
    \centering
    \includegraphics[width=\columnwidth]{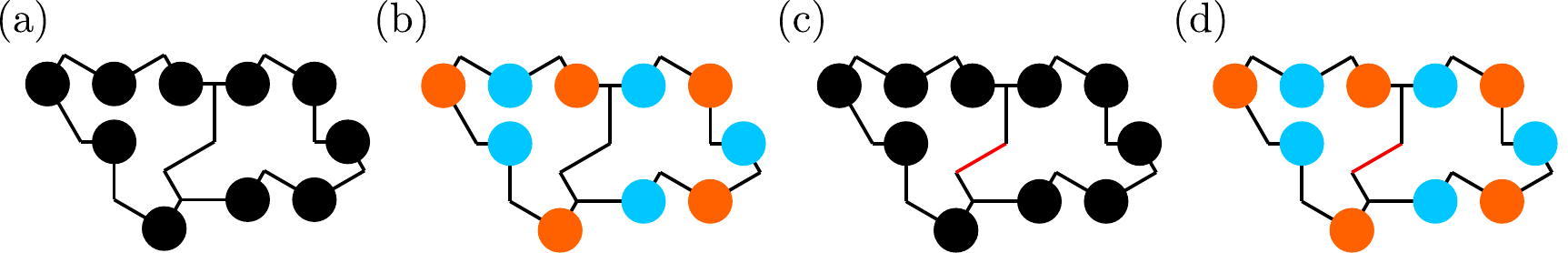}
    \caption{\label{fig:P0_zeromodes} {Properties of a zero-mode wave-function $\Psi_{zm}$ in case of the P0 Hat. }
Panels (a) and (b) show $|\Psi_{\rm zm}|^2$ and $\Psi_{\rm zm}$, respectively, in real-space for the case of $0$-flux.
Panels (c) and (d) show $|\Psi_{\rm zm}|^2$ and $\Psi_{\rm zm}$, respectively, in real-space for the case of $\pi$- flux. 
In all panels, black lines represent positive hoppings and red lines negative hoppings. }
\end{figure}

An exception to this observation are the P and F Hat tilings in generation $0$, that consist of the same arrangement of hat tiles. 
For this reason, we discuss only P0 Hat tiling in the following.
It has no anti-hats but hosts a single zero-mode.
This is surprising since none of the tiles in these Hats have $2m+2$ vertices, required to host a zero-mode in presence of a $\pi$-flux.
We observe that these zero-modes are localized at the outer edges of respective systems and their LDOS and $\Psi_{\rm zm}$ is identical to the LDOS and $\Psi_{\rm zm}$ of zero-modes in the case of $0$-flux, see Fig.~\ref{fig:P0_zeromodes}. 
This suggests that zero-modes in both cases ($0$- and $\pi$-flux) fulfill the same set of constraints.
We can understand this by noting that it is possible to find a gauge in which the $\pi$-flux is implemented via a single, negative hopping amplitude between vertices that are shared between hats that form P0 Hat tiling. 
One possible such configuration of hopping amplitudes is shown in Figs.~\ref{fig:P0_zeromodes}
(c-d).%
Therefore, the hopping amplitudes between vertices belonging to the boundaries of the sample can always be considered identical. 
Since there are $4m$ $(m=5)$ such vertices, the system can host a zero-mode under the same conditions it hosts a zero-mode in case of $0$-flux.

\begin{table}[h!]
\begin{tabular}{|l||ll|ll|ll|ll|}
\hline
\multicolumn{1}{|c|}{} & \multicolumn{2}{c|}{H}         & \multicolumn{2}{c|}{T}       & \multicolumn{2}{c|}{P}       & \multicolumn{2}{c|}{F}       \\ \hline
Inflation                      & \multicolumn{1}{l|}{AH}  & ZM  & \multicolumn{1}{l|}{AH} & ZM & \multicolumn{1}{l|}{AH} & ZM & \multicolumn{1}{l|}{AH} & ZM \\ \hline
0                              & \multicolumn{1}{l|}{1}   & 1   & \multicolumn{1}{l|}{0}  & 0  & \multicolumn{1}{l|}{0}  & 1  & \multicolumn{1}{l|}{0}  & 1  \\
1                              & \multicolumn{1}{l|}{3}   & 3   & \multicolumn{1}{l|}{1}  & 1  & \multicolumn{1}{l|}{2}  & 2  & \multicolumn{1}{l|}{2}  & 2  \\
2                              & \multicolumn{1}{l|}{22}  & 22  & \multicolumn{1}{l|}{3}  & 3  & \multicolumn{1}{l|}{12} & 12 & \multicolumn{1}{l|}{14} & 14 \\
3                              & \multicolumn{1}{l|}{147} & 147 & \multicolumn{1}{l|}{8}  & 8  & \multicolumn{1}{l|}{84} & 84 & \multicolumn{1}{l|}{98} & 98 \\ \hline
\end{tabular}
   \caption{Number of anti-hats (AH) and exact zero-modes (ZM) for each generation of inflation rules for magnetic flux $\phi/\phi_0=1/2$. }
    \label{tab:piBmodes}
\end{table}

Another exception we have observed is in quasicrystallites with anti-hats at the boundary (which are therefore not generated by inflation).
In these cases there are fewer zero-modes than anti-hats. 
In such cases the environment around the boundary anti-hats can frustrate the zero-mode.

\section{Zero-mode real-space structure and counting}
The electronic zero-modes in the tight-binding model are solutions to the equation
\begin{align}
H\ket{\psi_{\mathrm{zm}}}=0,
\end{align}
where $H$ is the tight-binding Hamiltonian described by Eq.~(1) in the main text.
This Hamiltonian defines the adjacency matrix of the undirected graph formed by the union of hats.
The number of zero-modes is then equal to the rank deficiency of $H$ (the number of columns minus the number of linearly independent columns).

\begin{figure*}
    \centering
    \includegraphics[width=1\textwidth]{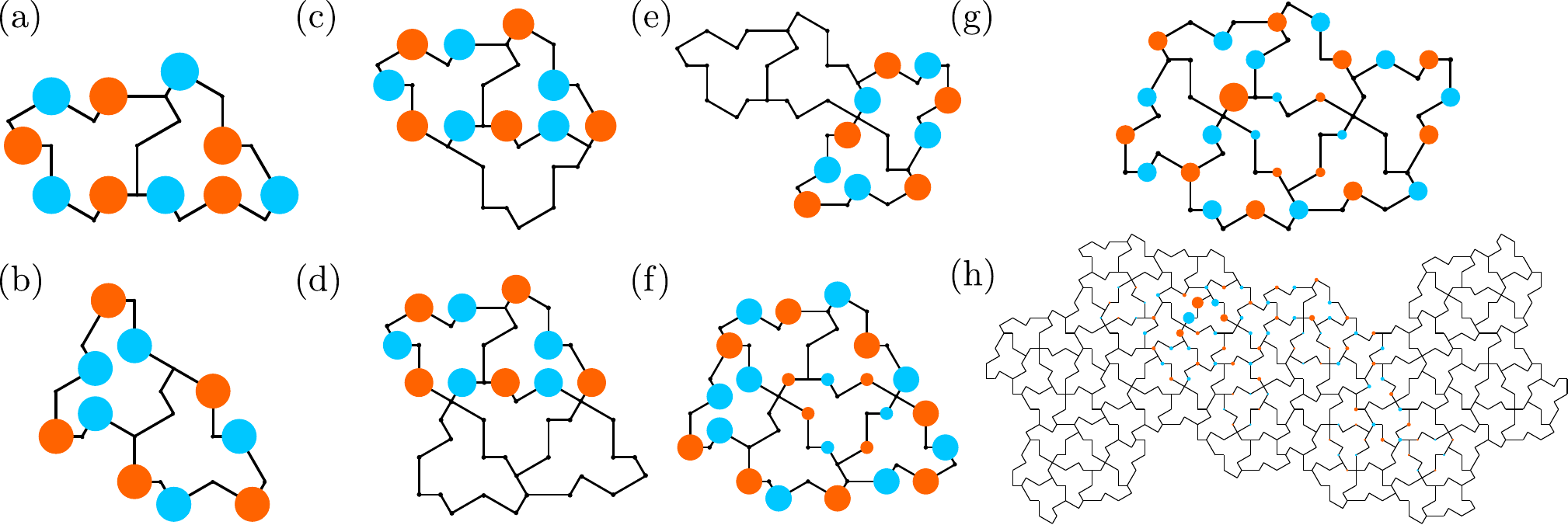}
    \caption{\label{fig:qcrystallites}{Amplitudes of the zero-mode wavefunction $\ket{\psi_{\mathrm{zm}}}$ for different quasicrystallites, including the P2 Hat tilling in panel (h).}
}
\end{figure*}

In the $0$-flux case it is possible to find a basis for the (un-normalized) zero-modes in which the amplitude on each vertex is an integer, and many vertices have zero amplitude. 
These are a generalization of the strictly localised zero-modes of Sutherland~\cite{Sutherland1986}. 
To illustrate this, we study several small quasicrystallites shown in Fig.~\ref{fig:qcrystallites}.
If a quasicrystallite system supports a zero-mode, we then plot the corresponding amplitudes of $\ket{\psi_{\mathrm{zm}}}$ in real-space.
From Figs.~\ref{fig:qcrystallites}(a-b), we see that whenever the graph contains a cycle of length $4m$, with $m$ integer, the amplitudes around the cycle can be taken to be the repeated sequence $\{0,1,0,-1\}^m$ and a system supports a zero-mode.
As illustrated in Figs.~\ref{fig:qcrystallites}(c-e), these zero-modes are preserved once additional hats are added to the quasicrystallite, provided that vertices from the rest of the graph connect only to cycle vertices of zero amplitude. 
This implies that the remaining graph needs to connect to vertices separated by even distances along the cycle, as these vertices can be chosen to have weight zero without implying nonzero weights off the cycle.

There may also be other integer-amplitude zero-modes not of the form just stated, see Figs.~\ref{fig:qcrystallites}(f-h). 
This is because, in general, it is possible to find integer-amplitude zero-modes by writing $H$ in its {Hermite normal form} $N$:
\begin{align}
N=UH,
\end{align}
where $N$ contains one column of zeroes for each zero-mode (there are various further restrictions on the form of $N$ which render the decomposition unique)~\cite{Kannan1979}. 
The integer unimodular matrix $U$ can be calculated efficiently using Gaussian elimination. 
For each zero column of $N$, the corresponding column in $U$ contains the amplitudes of the wave-function on each graph vertex. 
For example, the graph of P2 metatile shown in Fig.~\ref{fig:qcrystallites}(h) supports a zero-mode with wave-function amplitudes that range in magnitude from 0 to 6. 
Since there is only one mode, there is no possibility of decomposing it into a different basis with smaller amplitudes.

\end{document}